\documentclass{aa}
\usepackage{graphicx}
\begin{document}
\title{Optical Linear Polarimetry of Ultra Cool Dwarfs\thanks{Based on 
data collected at ESO/VLT with the FORS1 instrument during observing
programs 68-C.0171 and 69-C.0679.}}
%
\titlerunning{Polarimetry of field brown dwarfs}
\authorrunning{M\'enard et al.}
\author{
Fran\c cois M\'enard\inst{1}
\and Xavier Delfosse\inst{1}
\and Jean-Louis Monin\inst{1,2} 
}
\offprints{Fran\c cois M\'enard, \email{menard@obs.ujf-grenoble.fr}}
\institute{Laboratoire d'Astrophysique, Observatoire de Grenoble, 
CNRS/UJF UMR~5571, 414 rue de la piscine, BP 53, F-38041 Grenoble
cedex 9, France
\and
Institut Universitaire de France 
}
\date{Received: / Accepted: }
\abstract{
We present optical linear polarimetry of 8 ultra cool field dwarfs,
with spectral types ranging from M9 to L8. The linear polarisation, P,
of each dwarf we measured is P$<$0.2\%.  Three dwarfs have
polarisations compatible with zero, two are marginal detections, and
three have significant polarisation. Due to their small distance, an
insterstellar origin for the detected polarisation can be safely ruled
out. Our detections confirm that dust is present in the atmosphere of
these brown dwarfs and that the scattering geometry is not
symmetric. Possibilities for asymmetry include the dwarfs rotating
rapidly and being oblate, or the cloud coverage in the atmosphere
being inhomogenous.
}
\maketitle
\section{Introduction}

Recent sky surveys have uncovered large populations of objects cooler
than M dwarfs (e.g., Delfosse et al. 1997; Kirkpatrick et al. 1999).
Among them, the L dwarfs were the first to be identified (Mart\'{\i}n
et al. 1997). They cover a range of effective temperature between
$\sim$2200~K and $\sim$1400~K and are characterized by the
presence of condensates (i.e., grains) in their atmospheres. Of
particular interest, (spectro-)photometric monitoring revealed
variability (e.g., Gelino et al. 2002; Bailer-Jones 2002) that was
attributed to the presence of rapidly evolving clouds of particles
covering the photospheres not uniformly. Depending on the geometry,
light scattering by these clouds may yield net disk-integrated
polarisation.

However, there are various ways by which the light of a star can be
polarised. The presence of a magnetic field can induce polarisation by
Zeeman effect. Light scattering by inhomogeneous clouds or rapid
rotation leading to an elliptical photospheric disk can also, in
principle, yield net disk-integrated polarisation. In this paper we
explore, from an observational point of view, the linear polarisation
properties of field brown dwarfs.

We present the linear polarisation measurements of eight ultra cool
dwarfs (i.e., 1 very late M and 7 L dwarfs) obtained in the red, at
768nm, in a first step to constrain the dust distribution across the
photospheres of cool objects. The observations and results are
presented in \S2. In \S3, we present arguments regarding the origin of
the detected polarisation. In \S4 we discuss realistic photospheric
scattering geometries and propose observational tests. The behaviour
of the polarisation as a function of T$_{\rm eff}$ is presented in
\S5.

\section{Observations and results}
All the polarimetric data were obtained with the imaging polarimetry
mode of FORS1 attached to Melipal, UT3 of ESO's VLT facility located
atop Cerro Paranal, Chile.  FORS1 is mounted at the cassegrain focus
and provides a classical set-up for accurate dual beam imaging
polarimetry\footnote{Details of the imaging polarimetry mode of FORS1
can be found at {\tt http://www.eso.org/instruments/fors1}.}.

\begin{table*}[ht]
\caption{(Spectro-)Photometric data for the targets}
\begin{tabular}{lllllllll}
\hline
name & I & I-J & J-K & sp.ty. & $\pi$ & v$\sin{i}$ & EW(H$\alpha$) &
phot. var. \\ & & & & & mas & & & mag \\
\hline
LHS~102B                 &17.0(3)  & 3.70(3) & 1.90(3) & L5(3)  &104.7(7)
&32.5$\pm$2.5(11) & $<$4.0(4) & \\ 

2MASSW~J001544.7+351603  &17.3(8)  & 3.50(8) & 1.58(2) & L2(2)  & 54.2(10)
&                 & 2(2)      & 0.02?(13)  \\ 

2MASSW~J003615.9+182110  &16.11(6) & 3.67(6) & 1.41(2) & L3.5(5)&114.2(6)
&15(12)           & $<$0.5(2) & $<$0.01(13)\\ 

DENIS-P~J0255.0-4700     &17.14(4) & 3.66(4) & 1.62(4) & L8(4,2)  &159.2(10)
&40$\pm$10(11)    & $<$2(4)   &  \\ 


DENIS-P~J200048.4-752306 &15.88(1) & 3.23(1) & 1.19(1) & M9(8)& 58.5(9)
&                 &           & \\

DENIS-P~J203608.6-130638 &18.24(1) & 3.54(1) & 1.17(1) &L1-L2(8)& 31.3(9)
&                 &           & \\

DENIS-P~J205754.1-025229 &16.61(1) & 3.42(1) & 1.56(1) &L0-L1(8)& 55.7(9)
&                 &           & \\

2MASSW~J222443.8-015852  &18.02(6) & 3.97(6) & 2.03(2) & L4.5(2)& 88.1(6)
&                 & 1(2)      &  0.08(13)  \\ 
\hline
\end{tabular}
\label{tab:generaldata}
\noindent Note: (1)From DENIS survey, Delfosse et al. (2002), 
in preparation; (2) Kirkpatrick et al. (2000); (3) Goldman et
al. (1999); (4) Mart\'{\i}n et al. (1999); (5) Reid et al. (2000); (6) Dahn
et al. (2002); (7) Proper motion of LHS~102A, from Van Altena et
al. (1995); (8) Following the (I-J) vs. spectral type relations of
Dahn et al. (2002); (9) Following the (I-J) vs. M$_I$ relation of Dahn
et al. (2002); (10) Following the M$_J$ vs. spectral type relation of
Dahn et al. (2002); (11) Basri et al. (2000); (12) Schweitzer et
al. (2001); (13) Gelino et al. (2002)
\end{table*}

The observations were carried out during the periods 9-11 December
2001 and 16-19 May 2002. The Moon was set at the time of each
observation.  All data were collected with a broadband I$_{\rm
Bessel}$ filter centered on 768nm and 138nm wide (FWHM).  The
efficiency and stability of the instrument was checked and confirmed
on 6 occasions by measuring the highly polarised star Vela 1-95. All
our measurements fell well within 1-$\sigma$ (i.e., $<$0.08\%) of the
catalog value and no night-to-night efficiency corrections were
applied. We also checked for instrumental polarisation by measuring 3
nearby non-magnetic white dwarfs. With a typical 1-$\sigma$ error bar
of $\sigma_{\rm P}$=0.02\% all three measurements are compatible with
zero polarisation. In the following we will assume the imaging set-up
to be free of instrumental polarisation.

The data were detrended in a standard way with {\sc noao/iraf}. We
applied bias subtraction, cosmetic correction for deviant pixels, and
division by twilight flatfields. The flatfield frames were obtained
without the polarising optics. Due to a slight but systematic
variation of the bias level from night to night, i.e., an increase by
1-2 ADU every day, we obtained and used new sets of calibrations
frames every night for safety.

Except for LHS 102B (Goldman et al. 1999), we selected our targets
from the 2MASS and DENIS near-infrared sky surveys.  They are listed
in Table\ref{tab:generaldata} together with photometric and
spectroscopic information. Columns 2, 3, and 4 list the I-band
magnitudes and I-J and J-K colors respectively. Column 5 lists the
spectral types. A range is given when the spectral type is estimated
from near-infrared colors (except for the M9
DENIS-P~J200048.4-752306), all others are confirmed by
spectroscopy. When there is ambiguity, the spectral types are given in
the Kirkpatrick et al. (1999) system. The last four columns list the
annual parallax, $\pi$, the projected rotational velocities,
v$\sin{i}$, the H$\alpha$ equivalent widths and the photometric
variability, when available. References are listed in parenthesis next
to the data and refer to the notes at the bottom of the table.

%
%

Our results are presented in Table \ref{tab:poldata}. In order, the
columns list the abreviated target name, the date of observation, the
modified Julian date of the middle of the observation, and the
polarisation data, P, its associated error $\sigma_{\rm P}$, and the
position angle of the plane of vibration of the E-vector in the
equatorial coordinate system, when P/$\sigma_{\rm P}\sim$ 3.0 or more. The
modified Julian date (MJD) is related to the julian date (JD) by MJD =
JD - 2400000.5.

\begin{table}[htb]
\centering
\caption{I-band linear polarisation data}
\begin{tabular}{llllll}
\hline
name &  obs. date & MJD & P &$\sigma_{\rm P}$& $\theta$ \\
     &  d/m/y  &  & (\%) & (\%)& $\deg$\\
\hline
LHS~102B  &10/12/01&52253.050&0.105&0.036& 70.1\\ 
2M~J0015  & 9/12/01&52252.047&0.065&0.032& --- \\
2M~J0036  & 9/12/01&52252.089&0.199&0.028& 17.6\\
D~J0255   &11/12/01&52254.048&0.167&0.040& 80.6\\
D~J2000   &17/05/02&52411.394&0.083&0.017&122.9\\
D~J2036   &19/05/02&52413.391&0.122&0.042&170.4\\
D~J2057   &18/05/02&52412.398&0.044&0.023& --- \\
2M~J2224  & 8/05/01&52251.046&0.095&0.046& --- \\
\hline
\end{tabular}
\label{tab:poldata}
\end{table}
\section{The origin of the polarisation}
\subsection{An interstellar origin?}

The annual parallaxes listed in Tab. \ref{tab:generaldata} place all
the objects between 6pc and 32pc from the Sun.  Leroy (1993, 1999)
measured a sample of 1000 stars within $\sim$150pc from the Sun. Out
to a distance of 50pc no significant interstellar polarisation is
found and only 18 stars have P$\ge 0.1$\% in the distance range
between 60pc and 90pc of the Sun, from Hipparcos
distances. Furthermore, they are found in small and well defined
regions of the sky, away from our targets. An interstellar origin for
the linear polarisation presented here is therefore extremely unlikely
and we rule it out.  {\sl A consequence of this result is that the
frequency of intrinsically polarised brown dwarfs in our sample (i.e.,
from M9 to L8 dwarfs) appears extremely high, $\sim$ 50\%} (i.e., 3/8
(37\%), or 5/8 (62\%) if we include the marginal detections, see
\S4)\footnote{We choose to quote 50\% by taking the average of the
two, hence 4/8.}. For comparison, in nearby FGKM stars (dwarfs and
giants), the fractions are 2.5\%, 7\%, 5.5\%, and 11\% respectively in
the distance range 60-90pc from the Sun (Leroy 1999). Those fractions
go to zero for smaller distances.

\subsection{Possible mechanisms for intrinsic polarisation}

A possibility to produce intrinsic linear polarisation is via the
presence of magnetic field, either from Zeeman splitting of atomic or
molecular lines or synchrotron emission. The evidence for magnetic
field in L dwarfs is not clear yet. Observations show that the
H$_{\alpha}$ activity rapidly declines from mid-M to L dwarfs (Gizis
et al. 2000).  This may result from the high electrical resistivities
of their cool, hence mostly neutral, atmospheres (Mohanty et
al. 2002).  On the other hand, Berger (2002) detected high persistent
levels of radio emission in 3 late M and L dwarfs, including
2MASSW~J003615.9+182110, the most highly polarised source in our
sample. Magnetic fields in the range 10-1000G are deduced assuming
that the radio emission is coronal and that it peaks sharply at
8.5~GHz. Therefore, the (gyro-)synchrotron processes invoked will not
lead to significant polarisation in the optical, especially linear
polarisation.

Zeeman splitting of atomic lines is also unlikely the source of the
linear polarisation. For comparison, a sample of Ap stars with a
$\sim1$kG dipolar field at the surface show a maximum net linear
polarisation of order of a few times 0.01\% only (Leroy 1995). This
polarisation, produced by Zeeman splitting in saturated atomic lines,
is maximum where a large number of atomic lines are present in the
spectrum, i.e., in the blue for these stars. For ultra cool dwarfs,
atomic absorption does not dominate in the red where we made our
measurements. Wider molecular bands do, but these are complex, made of
numerous molecular lines side by side (e.g. Valenti et al. 1998) and
their global Zeeman polarisation, especially linear, is almost always much
lower than that of atomic lines.

Pending definitive measurements of the magnetic fields, we will assume
that they are not powerful enough at the surface of late-M and L type
brown dwarfs to induce a detectable linear polarisation in the
I-band. Scattering by photospheric dust grains therefore remains the
most likely mechanism for the polarisation.

\section{The scattering geometry: oblate photospheres or 
inhomogeneous dust clouds?}
\subsection{The polarised sources}
In a recent paper, Sengupta \& Krishan (2001, hereafter SK01) argued
that the photosphere of a brown dwarf will in general be oblate due to
fast rotation. The fast rotation is suggested by the large v$\sin{i}$
values measured in late-M and L dwarfs (Basri et al. 2000). This
oblateness will result in an asymmetric scattering geometry where
cancellation of the contribution of each point on the photosphere is
not perfect (as it would on a sphere) and a net disk-integrated linear
polarisation results. They calculated the linear polarisation expected
from single and multiple scattering by uniformly distributed dust in
an oblate photosphere seen edge-on (see their Figs. 1 \& 2
respectively).

In our sample, 3 targets have a significant polarisation:
2MASSW~J003615.9+182110, DENIS-P~J0255.0-4700, and
DENIS-P~J200048.4-752306. The maximum polarisation we detect is
$\sim$0.2\% at 768nm.

From these data, the single scattering case of SK01 for highly
eccentric ($e>0.3$) photospheres and most grain sizes except the very
smallest ($\alpha << 1.0$) can be excluded, because they would produce
too much polarisation, unless all targets in our sample are seen
pole-on and their projected photospheric disk is circular. This is
unlikely, and not compatible with the v$\sin{i}$ values presented in
Tab.\ref{tab:generaldata} for 2MASSW~J003615.9 and DENIS-P~J0255.

Multiple scattering usually lowers the polarisation because the planes
of the scattering events are randomly oriented and average each
other's contribution out from the final polarisation. At 768nm, the
predictions made by SK01 reflect that fact but the curves for small
(0.1$\mu$m) and large (1.0$\mu$m) grains are too close to the
polarisations we measured to choose between the two cases.
Measurements at longer wavelengths are needed to decide.

On the other hand, the models of SK01 do not consider the possibility
that the dust is distributed non uniformly in the photosphere. This
configuration may also be relevant to produce net linear polarisation
and in principle does not require projected photospheric oblateness.
Schubert \& Zhang (2000) argue that the dust in L dwarfs should be
organized in one of two states: in bands as in the giant planets of our
Solar System or in chaotic clouds.  

Although numerical simulations are needed to assess the polarising
power of these configurations, predictions can be made on geometrical
arguments. An oblate and uniformly dusty photosphere will always
produce stable polarisation, with both the position angle and the
polarisation level fixed, as the scattering geometry is constant. Dust
bands in the atmospheres will also produce a stable polarisation. On
the other hand, a photosphere covered with randomly distributed clouds
is likely to see its polarisation change because the scattering
geometry will change with time as the object rotates and the clouds
form, move and disappear. Therefore, the detection of variable linear
polarisation, especially variations of the polarisation position
angle, would clearly point toward cloud covered photospheres rather
than homogeneous or banded dust distributions in ultra cool dwarfs.

\subsection{The unpolarised sources}

In our sample, there are also three objects whose polarisation is not
detected, and two more which are just marginally detected, i.e., with
P/$\sigma_P$ at 2.9, just below 3.  Many possibilities exist to
explain their non detection. Obviously they could be devoid of dust,
but this is unlikely in view of the models and observations. For
example, 2MASSW~J222443.8-015852 is the only target in our list with a
confirmed photometric variability, presumably caused by inhomogeneous
dust clouds, but its polarisation is not detected.

Also, the objects that do not have v$\sin{i}$ measurements could be
slow rotators, or could be seen close to pole-on, hence showing a low
ellipticity to the observer and/or symmetric dust band structure.
Although statistically improbable, this possibility cannot be ruled
out yet.

To yield a low net polarisation in the optical, the scattering grains
may also be much larger than 1$\mu$m, which would agree with recent
theoretical predictions for dust size in L dwarf atmospheres. Also, it
is possible that the photospheres are covered by a very large number
of small randomly distributed dust clouds. Such a configuration should
not produce a large detectable polarisation. A few large clouds
probably being more favorable.

\section{Linear polarisation vs. spectral type}

Models predict that the amount and vertical location of dust is a
function of T$_{\rm eff}$ (e.g., Allard et al. 2001). In order to
match the rapid blueing of mid-L dwarfs, Ackerman \& Marley (2001)
suggest that the horizontal distribution of the dust is also a
function of T$_{\rm eff}$.

\begin{figure}
\includegraphics[width=5.0cm,angle=-90.0]{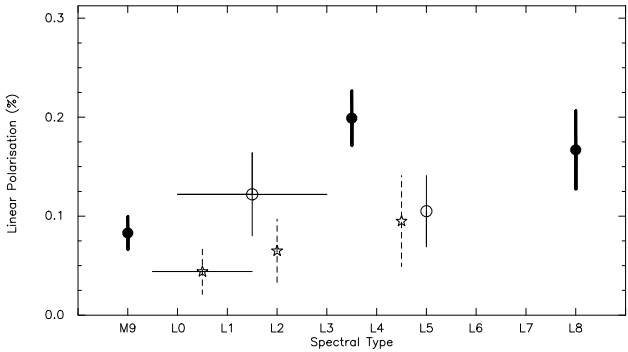}
\caption{\label{fig:polvsspt} Plot of the measured linear polarisation 
as a function of spectral type. The thick circles are detections.  The
marginal detections (i.e., P/$\sigma$(P) = 2.9) are the open
circles. The open stars with dashed error bars are the non-detections
(i.e., P/$\sigma$(P)$<$ 2.1). Stars with uncertain spectral types have
error bars plotted along the spectral type direction. 1-$\sigma$
polarisation error bars are plotted.}
\end{figure}

Fig.\ref{fig:polvsspt} is a plot of the measured linear polarisation
as a function of spectral type. Different symbols are used for
detections and non-detections (see the figure caption for details). A
slight trend for larger polarisation in cooler objects may be present,
but more data are clearly needed before we dare claim of anything
real. It would be interesting to extend the sample to include cooler
objects as their dust is expected to settle below the photosphere and
no polarisation from scattering should result.

\section{Conclusions}

We have measured the linear polarisation of one very late-M and seven
L dwarfs at a wavelength of 768nm. We have 3 detections, 2 marginal
cases, and 3 unpolarised targets.  In all cases the polarisations
remain low, below P=0.2\%. The linear polarisation is intrinsic to the
objects and not of interstellar origin. The fraction of polarised
nearby brown dwarfs in our sample is high, $\sim$50\%. It appears much
higher than for nearby FGKM stars.

Our small sample does not allow to identify the mechanism responsible
for the linear polarisation. However, fast spinning dwarfs with oblate
photospheres and uniform, or banded, dust clouds are expected to
produce a polarisation constant in time. On the other hand, large
randomly distributed dust clouds may produce more erratic
polarisations. Searching for polarimetric variability is needed to
solve this issue.

For now, we find no definite correlation of polarisation with spectral
type although we note a potential trend upward for cooler objects, up
to spectral type mid-L. More data are clearly needed, as well as the
extension of the sample to cooler T dwarfs. These objects are not
expected to have significant amounts of dust in their photospheres and
should therefore not show detectable linear polarisation.

\begin{acknowledgements}
It is a pleasure to thank Emanuela Pompei and Michael Dahlem from ESO
for their dedicated and efficient support during the observations. The
expertise of Thomas Szeifert with FORS1 polarimetry data is also
gratefully acknowledged. This research has made use of the SIMBAD
database, operated at CDS, Strasbourg, France. We thank the {\sl
Programme National de Physique Stellaire} (PNPS) for partial financial
support.
\end{acknowledgements}
\end{document}